\documentclass{ws-ijmpa}
\usepackage[super,compress]{cite}
\usepackage[pdftex,breaklinks]{hyperref}
\usepackage{breakurl}

\usepackage{graphics,graphicx}

\def\gsim{\mathrel{\rlap{\lower4pt\hbox{\hskip1pt$\sim$}}
    \raise1pt\hbox{$>$}}}         %greater than or approx. symbol
\def\lsim{\mathrel{\rlap{\lower4pt\hbox{\hskip1pt$\sim$}}
    \raise1pt\hbox{$<$}}}         %less than or approx. symbol

\newcommand{\bea}{\begin{eqnarray}}
\newcommand{\eea}{\end{eqnarray}}
\newcommand{\be}{\begin{eqnarray*}}
\newcommand{\ee}{\end{eqnarray*}}

\newcommand{\la}{\left\langle}
\newcommand{\ra}{\right\rangle}

\newcommand{\lp}{\left(}
\newcommand{\rp}{\right)}

\begin{document}

\markboth{Juan Rojo}
{Constraints on PDFs and $\alpha_s(M_Z)$ from LHC jet data}

%%%%%%%%%%%%%%%%%%%%% Publisher's Area please ignore %%%%%%%%%%%%%%%%
\catchline{}{}{}{}{}
%%%%%%%%%%%%%%%%%%%%%%%%%%%%%%%%%%%%%%%%%%%%%%%%%%%%

\title{CONSTRAINTS ON PARTON DISTRIBUTIONS AND
THE STRONG COUPLING FROM LHC JET DATA
}

\author{JUAN ROJO}
\address{Rudolf Peierls Centre for Theoretical Physics, 1 Keble Road,\\ University of Oxford, OX1 3NP Oxford, United Kingdom\\
Juan.Rojo@physics.ox.ac.uk}

\maketitle

\begin{history}
\received{Day Month Year}
\revised{Day Month Year}
\end{history}

\begin{abstract}
Jet production at hadron colliders provides powerful constraints on the
parton distribution functions (PDFs) of the proton, in particular
on the gluon PDF.
Jet production can also be used to extract the QCD coupling $\alpha_s(Q)$
and to test its running with
the momentum transfer up to the TeV region.
In this review, I summarize the information on
PDFs and  the strong coupling that has been provided
by Run I LHC jet data.
First of all, I discuss why jet production is directly sensitive
to the gluon and quark PDFs at large-$x$, and then
review the state-of-the-art perturbative calculations for
jet production at hadron colliders and the corresponding
fast calculations required for PDF fitting.
Then I present the results of various
recent studies on the impact on PDFs, in particular the gluon,
that have been performed using as input jet measurements
from ATLAS and CMS.
I also review the available determinations of the strong coupling constant 
based on ATLAS and CMS jet data, with emphasis
on the fact that LHC jet data provides, for the first time, a direct
test of the $\alpha_s(Q)$ running at the TeV scale.
I conclude with a brief outlook on possible 
future developments.

\end{abstract}

\ccode{PACS numbers: 13.87.-a, 12.38.Bx, 12.38.Qk}

%%%%%%%%%%%%%%%%%%%%%%%%%%%%%%%%%%%%
\section{Introduction}
\label{sec:Intro}
%%%%%%%%%%%%%%%%%%%%%%%%%%%%%%%%%%%%

The backbone of global fits of
parton distribution functions (PDFs) is provided
by deep-inelastic scattering (DIS)
measurements from fixed-target experiments and from the HERA collider.
While DIS data provides stringent constraints on the quark PDFs, it
is only indirectly sensitive to the gluon PDF through the
scaling violations encoded in the Dokshitzer-Gribov-Lipatov-Altarelli-Parisi (DGLAP)
evolution equations~\cite{dok,ap,gl}.
Thanks to the available lever arm, 
at not too small values of Bjorken-$x$ these scaling violations
and the precise inclusive HERA data\cite{Aaron:2009aa,Abramowicz:2015mha}
allow a reasonably accurate  determination of the gluon
PDF.
However, at medium and large values of  $x$,
the gluon is virtually unconstrained from DIS data, and thus
affected by large uncertainties.

For this reason, global PDF fits require complementary measurements
that are directly sensitive to the gluon PDF in the medium and
large-$x$ region.
When the first precision measurements of jet production
at Run I at the Tevatron\cite{Abe:1988as,Abe:1991ea,Abbott:1998ya} were
released, it became apparent that it should be possible to
access the large-$x$ gluon using differential distributions
in jet production.
In addition, it was recognized that a proper estimate of the PDF uncertainties
was essential for any search for New Physics at large
transverse energies ($E_T$) involving
jets in the final state.\cite{Martin:2004ir,Lai:1996mg}
With this motivation of constraining the poorly
known large-$x$ gluon, the latest inclusive jet measurements
 from the Tevatron Run II\cite{Aaltonen:2008eq,D0:2008hua} are included
in most global PDF fits.\cite{Ball:2012cx,Martin:2009iq,Gao:2013xoa}

While for quite some time Tevatron jet measurements were the only
ones available, since a few years a plethora of jet production
data at the Large Hadron Collider (LHC) is allowing us to test our
understanding of perturbative Quantum Chromodynamics (QCD) 
to an unprecedented level 
of precision, and to provide unique constraints on the gluon
and quark PDFs at large-$x$.\footnote{A detailed review of the constraints
  on PDFs obtained at Run I has been presented in the PDF4LHC
  report\cite{Rojo:2015acz}} 
Jet measurements from ATLAS and CMS are available from the 2010, 2011 and 2012
data taking periods of Run I, either in a final or
in a preliminary format, and several of these datasets have
already been used in PDF studies.
These measurements cover a wide kinematic range in jet transverse
momentum, dijet invariant mass, and rapidity, and provide precious information
on the large-$x$ PDFs.

In addition, the LHC jet data have made possible a number
of determinations of the strong coupling constant $\alpha_s$, 
and allowed for the first time to test its running at the TeV
scale.
The running of $\alpha_s$ is dictated
by the renormalization group (RG) evolution equations, and could
be modified as compared to the Standard Model (SM) predictions in the
presence of New Physics beyond the SM, for example
for new colored sectors.
While available determinations are typically limited by the scale
uncertainties from the next-to-leading order (NLO) calculations, recent progress
on the full next-to-next-to-leading order (NNLO) corrections suggest that such theory errors
will be substantially reduced in the near future.

In this review, I review the constraints that jet measurements
from the LHC have provided on the proton PDFs, specially
on the gluon, and on the
strong coupling constant.
The structure of this review is the following.
I begin in Sect.~\ref{sec:sensitivity} with a discussion
of the PDF sensitivity of jet production.
Then in Sect.~\ref{sec:theory} I review the theoretical
calculations and tools that allow us to extract information
on PDFs and $\alpha_s$ from the collider jet data.
I continue in Sect.~\ref{sec:constraints} presenting
available studies that quantify the impact of
LHC jet data on the PDFs, and in Sect.~\ref{sec:alphas}
I summarize available determinations of the strong coupling
from LHC jet production.
I conclude in Sect.~\ref{sec:outlook} and discuss the outlook
for future measurements and studies based on the LHC jet data.

Let us recall also that jet production is not the only final state
through which the gluon PDF and $\alpha_s$ can be probed at the LHC.
Related complementary 
processes include isolated photon production\cite{d'Enterria:2012yj}, 
top quark
pair production\cite{Czakon:2013tha,Chatrchyan:2013haa} and 
high-$p_T$ $Z$ boson production\cite{Malik:2013kba}.
In particular, top quark pair production benefits from
reduced theory uncertainties from the recent NNLO calculation\cite{Czakon:2013goa},
though it cannot compete with jet production in terms
of kinematical reach for the PDFs.
For the small-$x$ gluon, it has been recently showed
how important information can be obtained
from the LHCb charm
production data~\cite{Gauld:2015yia,Zenaiev:2015rfa}.

%%%%%%%%%%%%%%%%%%%%%%%%%%%%%%%%%%%%%%%%%%%%%%%%%%%%%%%%%%
\section{PDF sensitivity of jet production in hadronic collisions}
%%%%%%%%%%%%%%%%%%%%%%%%%%%%%%%%%%%%%%%%%%%%%%%%%%%%%%%%%%
\label{sec:sensitivity}

\begin{figure}[t]
\begin{center}
\includegraphics[width=0.95\textwidth]{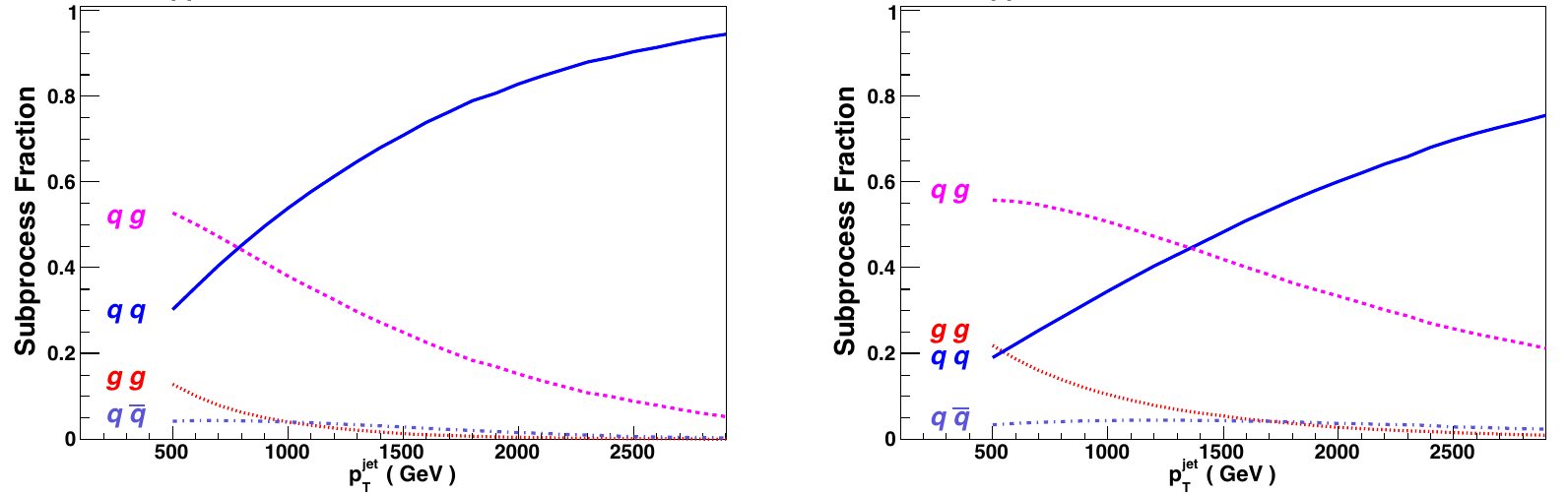}
\end{center}
\vspace{-0.3cm}
\caption{\label{fig:jetsubprocess} Subprocess decomposition of
inclusive jet production at the LHC 8 TeV (left plot) and
14 TeV (right plot), computed with ALPGEN at LO using MSTW08
as input PDFs, as a function of the $p_T$ of the leading jet.
}
\end{figure}

To quantify the PDF coverage of LHC jet data,
it is useful to review the kinematics of jet production at hadron colliders.
For simplicity, one can work in the Born 
approximation, where the leading processes are of the form 
\be
{\rm parton}_i(p_1)+{\rm parton}_j(p_2) \, \to \,{\rm parton}_k(p_3)+{\rm parton}_l(p_4) 
\, ,
\ee
with $i,j,k,l$ flavor indices to denote either quarks, antiquarks
or gluons.
In Fig.~\ref{fig:jetsubprocess} I show the subprocess decomposition of
inclusive jet production at the LHC 8 TeV (left plot) and
14 TeV (right plot), computed with {\sc\small ALPGEN}\cite{Mangano:2002ea} at 
leading order using MSTW08\cite{Martin:2009iq} as input PDFs.

From Fig.~\ref{fig:jetsubprocess} we see that at 
8 TeV $qg$ scattering is the main production mechanism for
$p_T^{\rm jet}\le$ 800 GeV, then $qq$ scattering becomes more important due
to the steeper fall-off of the gluon PDF at large-$x$.
At the LHC 14 TeV, gluon-initiated contributions dominate
up to $p_T^{\rm jet}\sim 1.5$ TeV.
Therefore, it is clear that there is a wide range in jet $p_T$ for
which gluon-initiated contributions dominate the cross-section, and
thus measurements in this range allow to pin down the gluon PDF.
At the highest values of $p_T$, jet production instead probes the large-$x$
quarks.
Note also that the contributions initiated by antiquarks are much smaller
since their PDFs are strongly suppressed as compared
to those of quarks and gluons at large-$x$.

%%%%%%%%%%%%%%%%%%%
%%%%%%%%%%%%%%%%%%%%%%%

Denoting by $y_3$ and $y_4$ the rapidities of the two outgoing
partons in the laboratory reference frame, and with $p_T$ being their
(back-to-back) transverse momentum, one finds that dijet production
probes the PDFs at the following values of their momentum fractions:
\be
x_1 \, = \, \frac{p_T}{\sqrt{s}}\,\lp e^{y_3}+e^{y_4}\rp \, , \quad
x_2 \, = \, \frac{p_T}{\sqrt{s}}\,\lp e^{-y_3}+e^{-y_4}\rp \, ,
\label{eq:born}
\ee
with $\sqrt{s}$ being the center-of-mass energy of the hadronic collision. 
Note
that in inclusive jet production instead the
underlying values of $x_1$ and $x_2$
are not uniquely defined in terms of the measurement final state
kinematics, usually taken to be $p_T^{\rm jet}$ and $y_{\rm jet}$.
Dijet production is typically measured as a function of the dijet
invariant mass $m_{34}$, which in the Born kinematics is given by
\be
m_{34} = 2\, p_T\,\cosh \lp \frac{y_3-y_4}{2} \rp \equiv 
 2\, p_T\,\cosh y^* \, ,
\ee
which is manifestly invariant under longitudinal boosts,
and where we have defined the rapidity difference of the two
outgoing partons as  $y^* \equiv (y_3-y_4)/2$.
Dijet production measurements are typically binned in $m_{34}$
and in $y^*$, and an upper limit on the maximum rapidity of the individual
jets $|y_{3,4}|\le y_{\rm max}$ is imposed.
It is thus possible to derive the range
of Bjorken-$x$ probed in dijet production, in terms 
of the measured final state kinematics, which is given by
\be
x_{\rm min}= \frac{m_{34}}{\sqrt{s}}e^{-y_{\rm max}+y^*} \le \, x \, \le 1 \,.
\label{fig:xmin}
\ee
Therefore in dijet production we
can have access to PDFs with Bjorken-$x$ from
$x_{\rm min}$ to one.
Similar expressions can be derived for inclusive jet production measurements.

%%%%%%%%%%%%%%%%%%%%%%%%%%%%%%%%%%%%%%%%%%%%%
\begin{figure}[t]
\begin{center}
\includegraphics[width=0.58\textwidth]{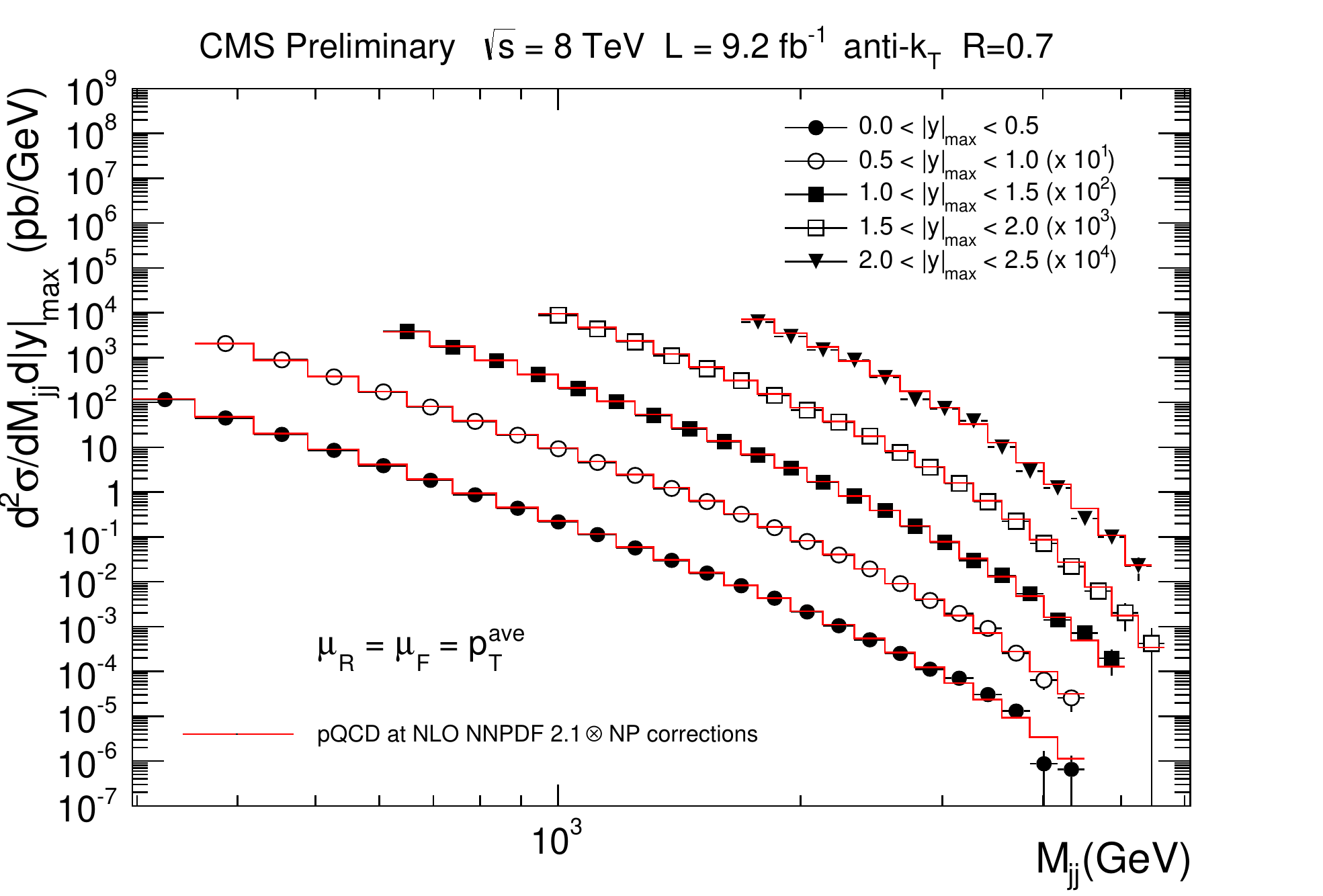}
\includegraphics[width=0.40\textwidth]{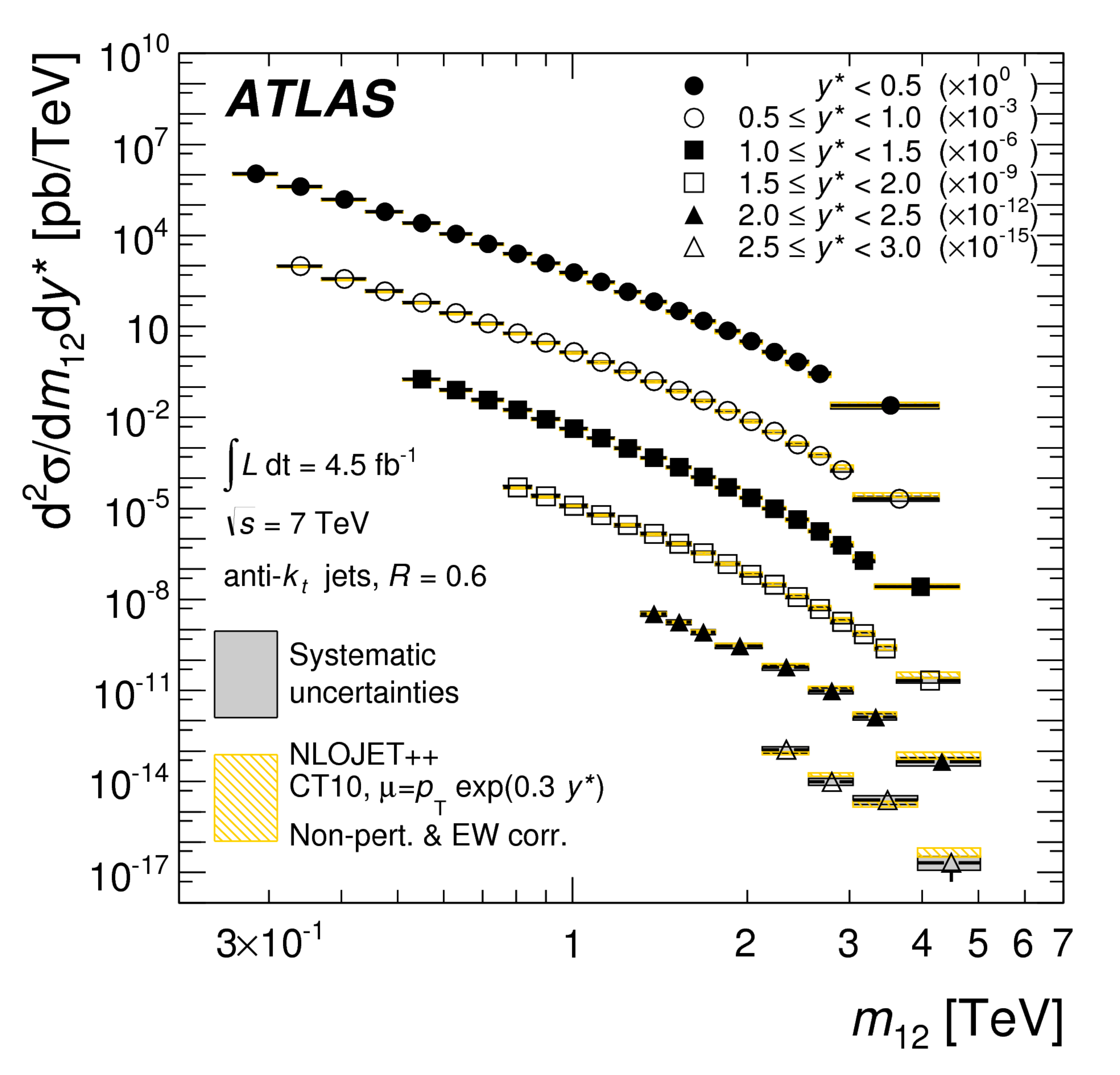}
\end{center}
\vspace{-0.3cm}
\caption{\label{fig:dijet} Left plot: preliminary CMS 
data\cite{CMS-PAS-SMP-14-002}
for dijet production at 8 TeV.
The measured cross-sections are compared to NLO QCD theory,
 using NNPDF2.1 as input PDFs.
Right plot: results for the ATLAS 2011 
dijet measurement\cite{Aad:2013tea}, now using CT10 as
input PDF.
In both cases the theoretical calculations include non-perturbative
and electroweak corrections.
}
\end{figure}
%%%%%%%%%%%%%%%%%%%%%%%%%%%%%%%%%%%%%%%%%%%%%

%%%%%%%%%%%%%%%%%%%%%%%%%%%%%%%%%%%%%%%%%%%
\begin{figure}[t]
\begin{center}
\includegraphics[width=0.49\textwidth]{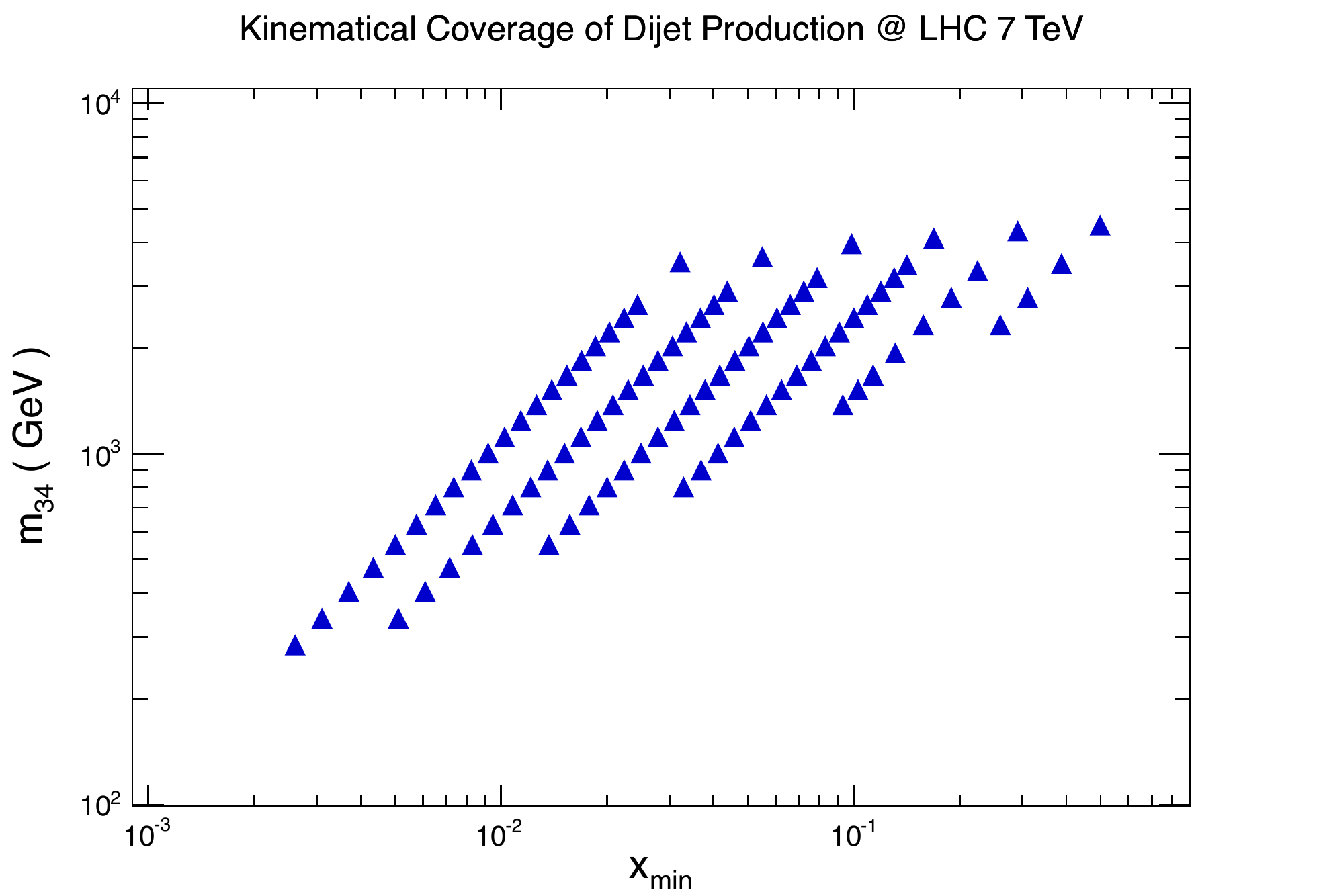}
\end{center}
\vspace{-0.3cm}
\caption{\label{fig:dijet2} Minimum value of Bjorken-$x$ and the
scale $m_{34}$ probed in the PDFs for dijet production at the LHC
7 TeV, using the kinematics of the ATLAS 2011 
dijet measurement\cite{Aad:2013tea}.
}
\end{figure}
%%%%%%%%%%%%%%%%%%%%%%%%%%%%%%%%%%%%%%%%%%%%%

As illustrative examples of the kinematical coverage
of LHC jet data, in Fig.~\ref{fig:dijet} (left plot) I show the 
preliminary CMS data\cite{CMS-PAS-SMP-14-002}
on dijet production at 8 TeV, where the results
 are compared to NLO QCD theory
 using NNPDF2.1\cite{Ball:2011mu} as input PDFs.
Note that the reach in the dijet invariant mass $m_{34}$ is almost 6 TeV.
Then in Fig.~\ref{fig:dijet2}
I show the value of the $x_{\rm min}$ probed
in the PDFs for dijet production at
7 TeV, using the kinematics of the  ATLAS dijet measurement
from the 2011 dataset\cite{Aad:2013tea}.
The corresponding cross-section measurements are shown in the 
right plot of Fig.~\ref{fig:dijet}.
As can be seen, in this particular case the dijet data probes the PDFs in the range $x \gsim 2\cdot 10^{-3}$
and for momentum transfers in
the range of $2\cdot 10^2 \lsim Q \lsim 5\cdot 10^{4}$ GeV.
In addition, the higher the invariant mass of the dijet system
$m_{34}$, the larger the value of Bjorken-$x$ that will be probed.

While Fig.~\ref{fig:dijet2} determines the  region of Bjorken-$x$ that
is kinematically accessible in jet production measurements,
it does not provide information on which part of this region
dominates the
production cross-section, or in other words, the region of
Bjorken-$x$ for which the PDF sensitivity of the jet data
is maximized.
To determine this important information, it is
possible to compute the correlation coefficients between
the PDFs and the experimental data.
As explained in Ref.\cite{Demartin:2010er}, in a Monte Carlo PDF set one
can compute the correlation between the parton
distributions, for different values of $x$ and $Q^2$, and the jet production
cross-sections, for different bins of jet transverse momentum
and rapidity.

\begin{figure}[t]
\begin{center}
\includegraphics[width=0.49\textwidth]{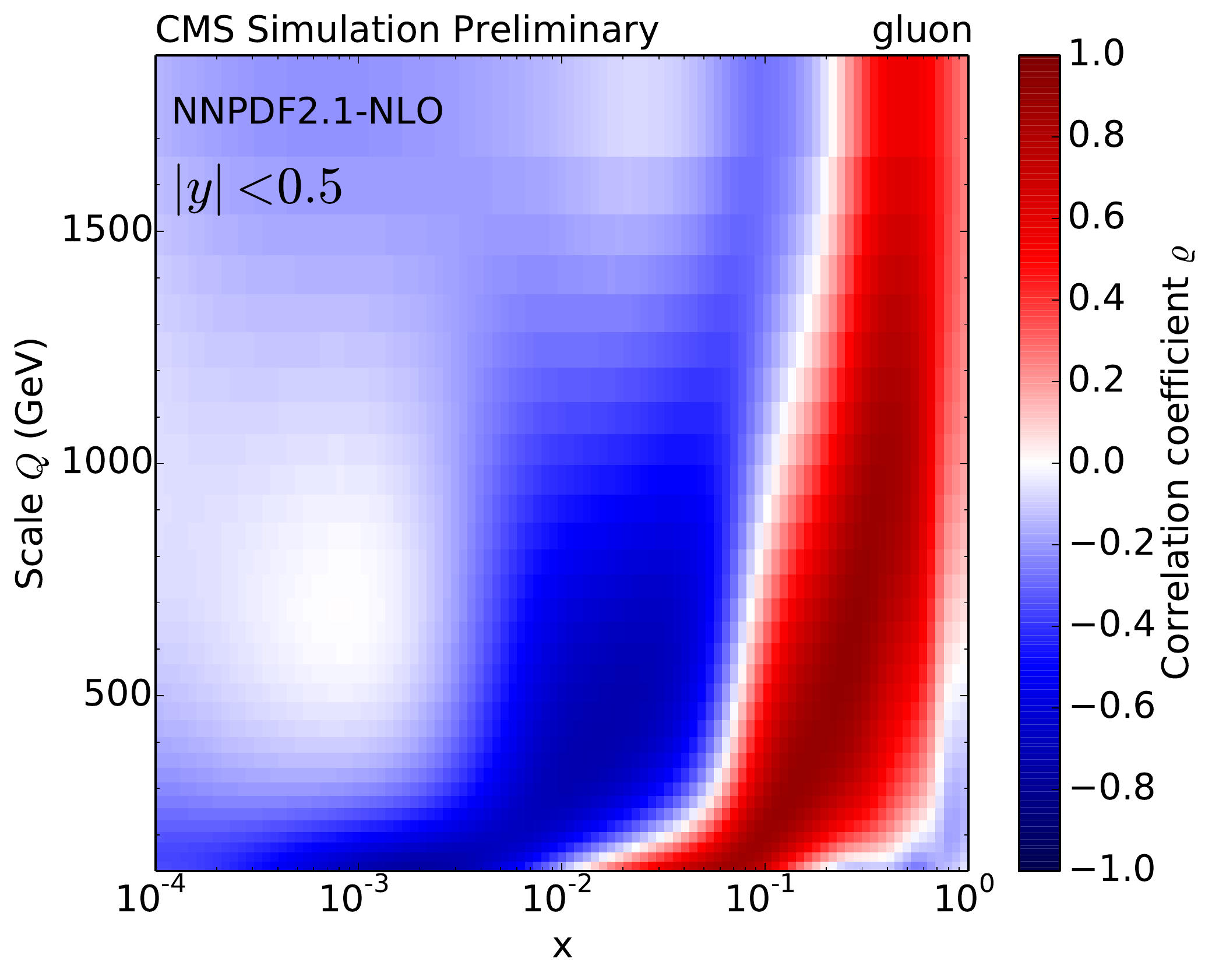}
\includegraphics[width=0.49\textwidth]{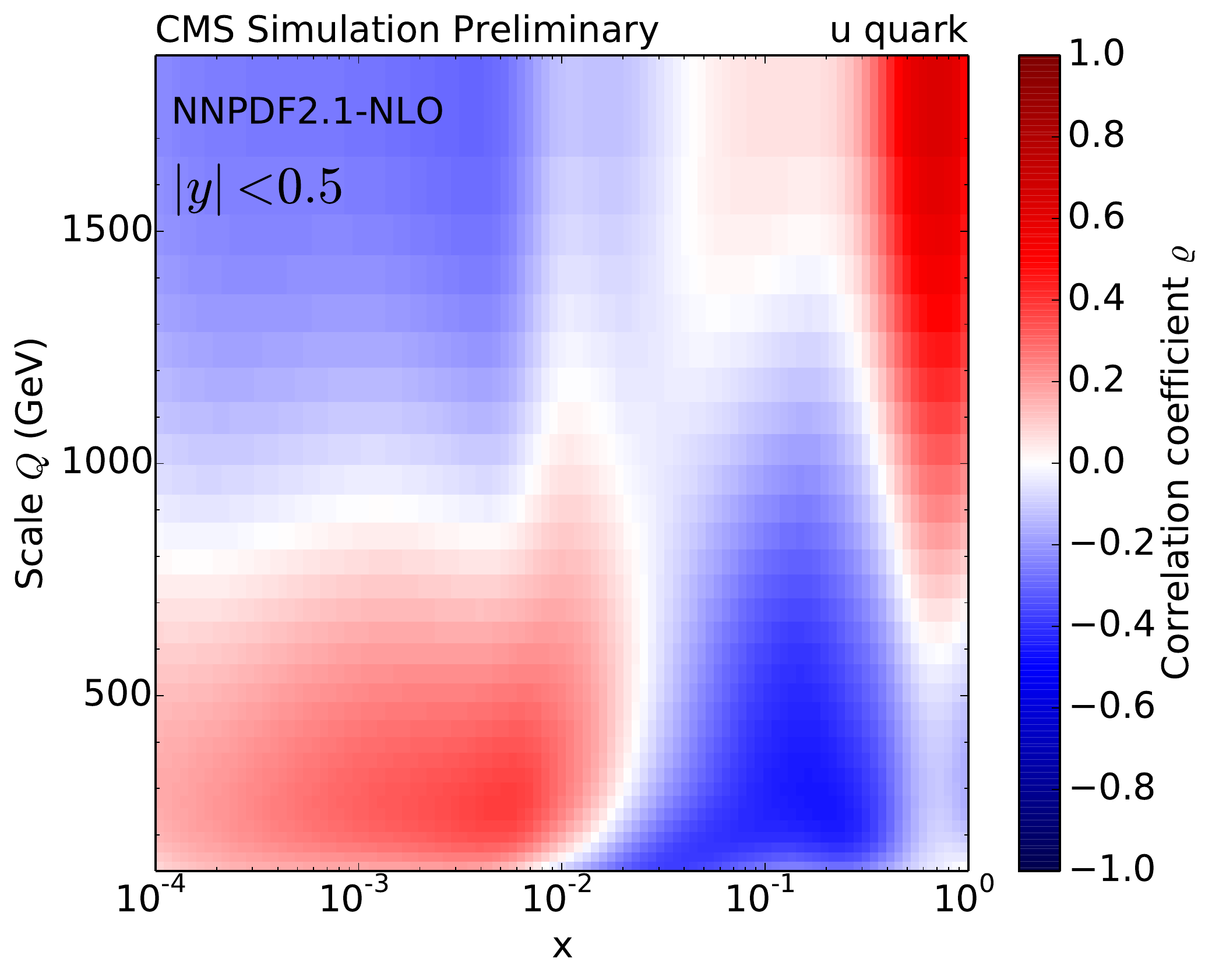}
\end{center}
\vspace{-0.3cm}
\caption{\label{fig:correlation} Correlation coefficients between the gluon PDF
(left plot) and the up quark PDF (right plot) for different values of
$\lp x,Q^2\rp$ for the kinematics of the CMS 7 TeV inclusive jet
cross-section.
The computation of these coefficients, performed with NNPDF2.1 NLO,
has been performed in all the jet $p_T$ bins
of the central rapidity region, $|y|\le 0.5$.
Results taken from Ref.\cite{Khachatryan:2014waa}, 
additional figures are available from \url{https://twiki.cern.ch/twiki/bin/view/CMSPublic/PhysicsResultsSMP12028}.
}
\end{figure}

Using NNPDF2.1 NLO, this exercise was carried out in the CMS analysis of 
Ref.\cite{Khachatryan:2014waa}, which studies the constraints on
PDFs and on $\alpha_s$ of their
7 TeV inclusive jet data.
The results can be found in Fig.~\ref{fig:correlation}, which shows
the correlation coefficient between PDFs (in this case the gluon and the
up quark) for all the $p_T$ bins in the central rapidity region, $|y|\le 0.5$, as a function of
Bjorken-$x$ and the momentum transfer $Q$.
A value of this coefficient close to one (minus one) indicates that, for this specific
data bin, the cross-section is strongly (anti-)correlated with the
corresponding PDFs in the given range of $x$.
In particular, from Fig.~\ref{fig:correlation} one can 
see that LHC inclusive jet data
has a strong correlation with the gluon for $x \ge 0.1$, with
a likewise strong anti-correlation for $x \sim 10^{-2}$.
This correlation is weaker for the up quark, except for
large values of $x$, that is, $x\gsim 0.4-0.5$, for which the 
$qq$ scattering channel begins to dominate over $qg$ scattering,
see Fig.~\ref{fig:jetsubprocess}.

%%%%%%%%%%%%%%%%%%%
%%%%%%%%%%%%%%%%%%%%%%%

%%%%%%%%%%%%%%%%%%%%%%%%%%%%%%%%%%%%
\section{Theory calculations and tools for fitting jet data}
%%%%%%%%%%%%%%%%%%%%%%%%%%%%%%%%%%%%
\label{sec:theory}

The NLO cross-sections for jet production at hadron colliders
have been known for a long time\cite{Ellis:1990ek,Nagy:2003tz}.
They have been implemented in various computer programs,
such as {\sc NLOjet++}\cite{Nagy:2001fj}.
Computing differential distributions for jet observables with
these codes is however very CPU-time intensive, and thus not suitable
for the aims of PDF determinations, where the iterative fitting procedure
requires recomputing the same observables a large number of
times.
With this motivation,
different fast interfaces to NLO jet calculations have been developed.
The basic idea of these interfaces is to
precompute the partonic cross-section in a grid in $\lp x_1,x_2,Q^2\rp$,
so that it is possible to perform 
an a posteriori convolution with the PDFs to yield the hadronic cross-section
much more efficiently.
In particular, the
{\sc NLOjet++} calculations have been interfaced to both {\sc APPLgrid}\cite{Carli:2010rw} 
and {\sc FastNLO}\cite{Wobisch:2011ij}, making possible to include
collider jet data into PDF fits without any K-factor approximation.
More recently, the {\sc\small aMCfast} interface\cite{Bertone:2014zva} to 
{\sc\small MadGraph5\_aMC@NLO}\cite{Alwall:2014hca} has been released.
{\sc\small aMCfast} also allows
to include NLO jet calculations into PDF fits, with the additional
possibility of accounting also for the matching of the fixed-order 
calculation to
parton showers.

Going beyond NLO, thanks to recent breakthroughs in our ability to perform
computations at NNLO in QCD for processes that include
colored particles both in the initial and final state\cite{GehrmannDeRidder:2005cm}, 
the full NNLO cross-section for inclusive jet and dijet production
in the $gg$ channel has recently become available\cite{Ridder:2013mf,Currie:2013dwa}.
This result is an important milestone 
towards the full NNLO calculation.
The ongoing work in the remaining partonic channels suggests that the full
NNLO cross-section for hadronic jet production will be available
in the near future.
These results are of paramount importance to extend the physics potential
of the interpretation of LHC jet measurements, often limited by the
theoretical systematics from the NLO scale variations.

While the full NNLO calculation of jet production becomes available,
it is possible to include jet data in NNLO fits
by using approximate NNLO calculations based on threshold
resummation, such as those derived in Refs.\cite{deFlorian:2013qia,Kidonakis:2000gi,Kumar:2013hia} .
The most recent of these calculations\cite{deFlorian:2013qia}, among
other improvements, includes the full dependence on the jet radius $R$
in the resummed result.
However, before being able to use these approximate NNLO calculations in
a PDF fit, it is crucially important to determine the range
of validity of the threshold approximation.
This can be done by comparing the  exact $gg$ NNLO calculation
with the approximate NNLO in the same partonic channel, using
identical cuts and binning as those
of the corresponding Tevatron and LHC jet measurements.

This idea has been exploited in Ref.\cite{Carrazza:2014hra}, which provides
the complete list of approximate NNLO K-factors, and their
region of validity, for all published Tevatron and LHC inclusive
jet production measurements.
This is all the information
that is needed in order to include
jet data into a NNLO fit. 
In particular, Ref.\cite{Carrazza:2014hra} finds that while the threshold
approximation is good at high $p_T$ and central rapidities, it is much poorer
at low $p_T$ and forward rapidities.
Therefore, it is possible to include most of the available LHC jet
measurements in a global NNLO PDF fit provided that one restricts the
fitted data to the central and high-$p_T$ regions.
This is the strategy that has been used  in the NNPDF3.0 fits\cite{Ball:2014uwa}.

Finally, one should mention that 
at the high transverse momenta and invariant masses that the LHC is and
will be covering,
electroweak corrections to jet production are non negligible.
For example, the calculation of purely weak radiative corrections for dijet
production at hadron colliders\cite{Dittmaier:2012kx}
finds corrections that can be as large as 10\% at the highest possible
jet transverse momenta.
While accounting for these effects is probably not essential for PDF fits to Run I jet
data, given the experimental
uncertainties in the high $p_T$ region, their inclusion will 
become mandatory with the Run-II data, which will explore deep into the TeV region.

%%%%%%%%%%%%%%%%%%%%%%%%%%%%%%%%%%%%
\section{PDF constraints from LHC jet data}
%%%%%%%%%%%%%%%%%%%%%%%%%%%%%%%%%%%%
\label{sec:constraints}

Using the theoretical calculations and the fast interfaces
that I have discussed in the previous section, it becomes possible
to include LHC jet data into PDF fits and quantify the constraints
that it provides.
First of all, I present the PDF studies that have been performed by the
ATLAS and CMS collaborations.
From the ATLAS side, the available measurements relevant for PDF
studies are the inclusive jet and dijet
cross sections from the 2010 dataset\cite{Aad:2011fc}, 
the ratio of 2.76 TeV to
7 TeV inclusive jet cross-sections\cite{Chatrchyan:2013txa}  and 
the dijet cross-sections
from the 2011 dataset\cite{Aad:2013tea}.

\begin{figure}[t]
\begin{center}
\includegraphics[width=0.49\textwidth]{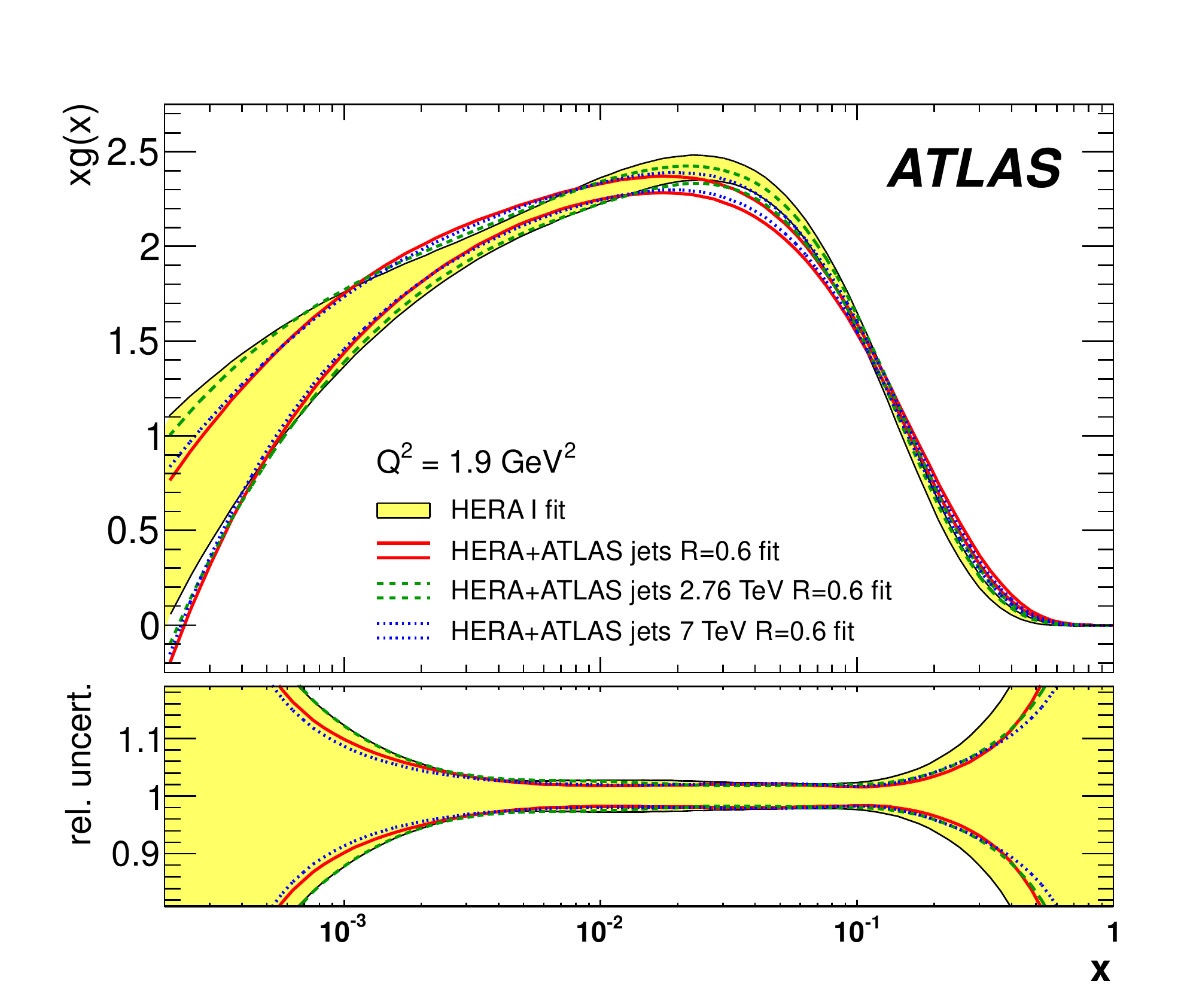}
\includegraphics[width=0.49\textwidth]{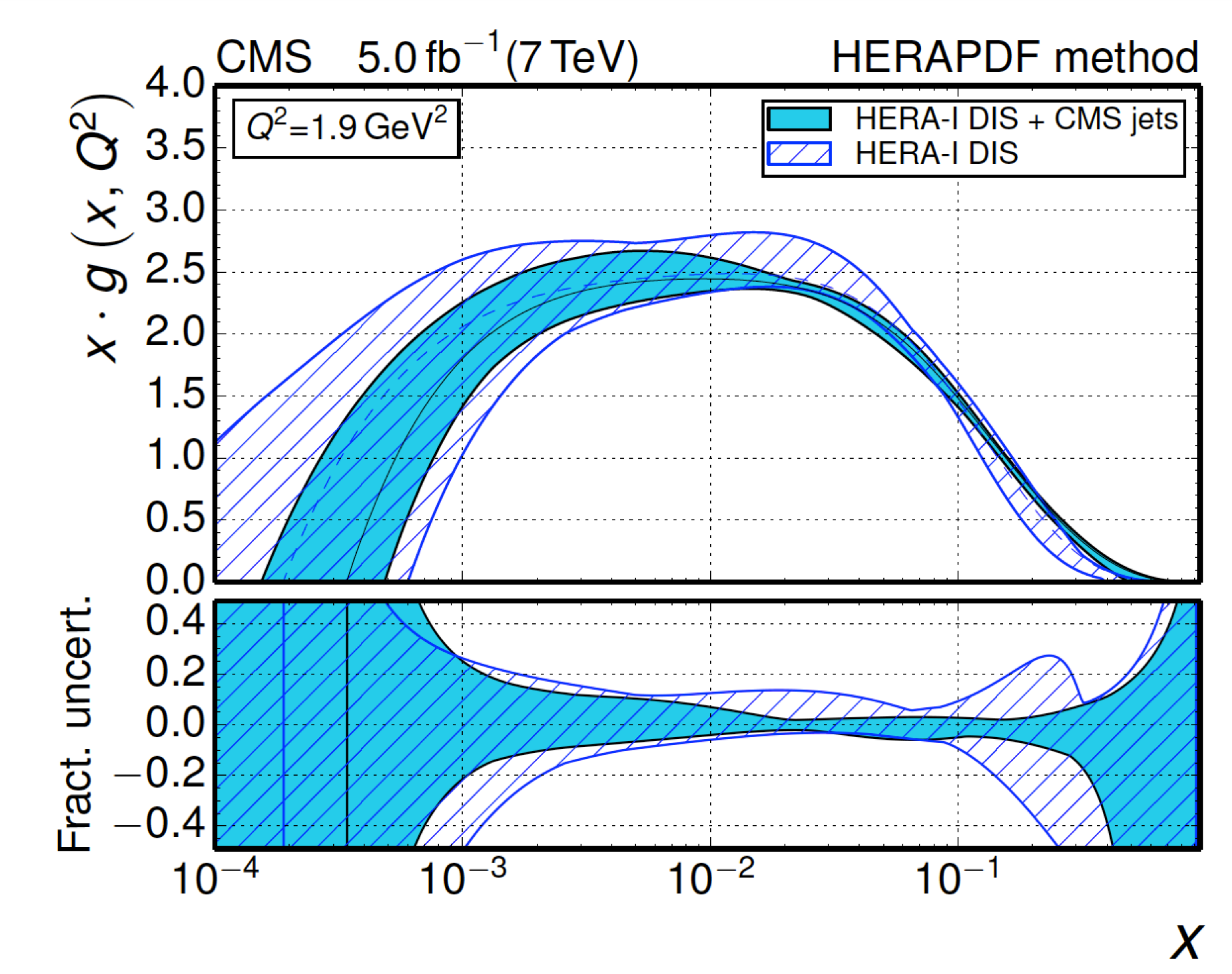}
\end{center}
\vspace{-0.3cm}
\caption{\label{fig:atlaspdfs} Representative results of
PDF analyses using jet data obtained by the ATLAS and CMS
collaborations. Left plot: impact on the gluon PDF of
the ATLAS data on the 2.76/7 TeV inclusive jet cross-section ratio\cite{Chatrchyan:2013txa}.
Right plot: constraints on the gluon from the CMS
inclusive jet 2011 data\cite{Khachatryan:2014waa}.
In both cases the results of the PDF fit are compared to a HERA-only
baseline fit.
}
\end{figure}

The first ATLAS QCD analysis of jet data was performed
using their measurement ratio of jet cross-sections between two different
center-of-mass energies.
The motivation for such a ratio measurement is that due
to the cancellation of several theoretical (in particular
scale variations) and experimental (jet energy scale) uncertainties,
there is a complementary PDF sensitivity as compared to the absolute
cross-section measurements\cite{Mangano:2012mh}.
In Ref.\cite{Chatrchyan:2013txa}, this PDF analysis of the ratio of
2.76 TeV over 7 TeV inclusive jet data was performed to demonstrate the
sensitivity to the large-$x$ gluon PDF, by comparing
to a baseline fit based on HERA data only.
The results of this analysis, based on the {\sc\small HERAfitter} framework,
are shown in Fig.~\ref{fig:atlaspdfs}.
The reduction on the gluon PDF large-$x$ uncertainties can be appreciated.

More stringent constraints can be
provided by the ATLAS jet measurements from the 2011
run.
The inclusive jet cross-section from the 2011 run has been
presented
in Ref.\cite{Aad:2014vwa}\,, where it is compared to
several modern PDF sets.
The ATLAS dijet analysis from the 2011 data\cite{Aad:2013tea}
contains a quantitative $\chi^2$ comparison for the predictions of different
PDF sets, showing a clear discrimination power.
In addition, the ATLAS measurement of
 three-jet production cross-sections~\cite{Aad:2014rma} from 
the 2011 data can also be potentially useful for PDF studies.
These measurements are particularly interesting for PDF fits since for the
first time the full correlation matrix between inclusive jet, dijet and three-jet
data is publicly available,
allowing to include in a PDF fit the complete ATLAS
jet production dataset from 2011.

%%%%%%%%%%%%%%%%%%%
%%%%%%%%%%%%%%%%%%%%%%%

CMS has also studied the PDF sensitivity of their jet production 
measurements.
The quantification of the PDF constraints from the 
inclusive jet cross sections measured in the 2011 dataset~\cite{Chatrchyan:2012bja} has
been studied by CMS in Ref.\cite{Khachatryan:2014waa}.
As expected, a substantial reduction on the medium and large-$x$ gluon
PDF uncertainties as compared to a baseline HERA-only fit is found,
as illustrated in Fig.~\ref{fig:atlaspdfs}.
This CMS analysis also emphasizes the crucial role of a careful estimation
of systematic uncertainties and their correlation to improve the PDF
sensitivity of jet measurements.
In this CMS study the traditional analysis based on 
{\sc\small HERAfitter} is complemented with a Monte Carlo
analysis with data-based regularization.
Ongoing measurements of inclusive jets\cite{CMS-PAS-SMP-12-012} 
and dijets\cite{CMS-PAS-SMP-14-002} at 8 TeV will
further extend the constraining power of the CMS measurements.

While not directly usable in PDF fits, the CMS measurement of the
ratio of jet cross-sections at different values of the jet
radius,  $R=0.5$ and $R=0.7$, see Ref.\cite{Chatrchyan:2014gia},
provides useful information on the theory systematics that affect these
measurements specially at moderate $p_T$\cite{Cacciari:2008gd,Dasgupta:2007wa}.
In this respect, note that ATLAS jet measurements  are provided for
two different jet radii, $R=0.4$ and $R=0.6$, see for example Ref.~\cite{Aad:2011fc},
allowing similar studies as those of the CMS paper, and in particular
providing an important cross-check that the impact of jet data on PDFs
does not depend strongly on the specific value of $R$.

\begin{figure}[t]
\begin{center}
\includegraphics[width=0.47\textwidth]{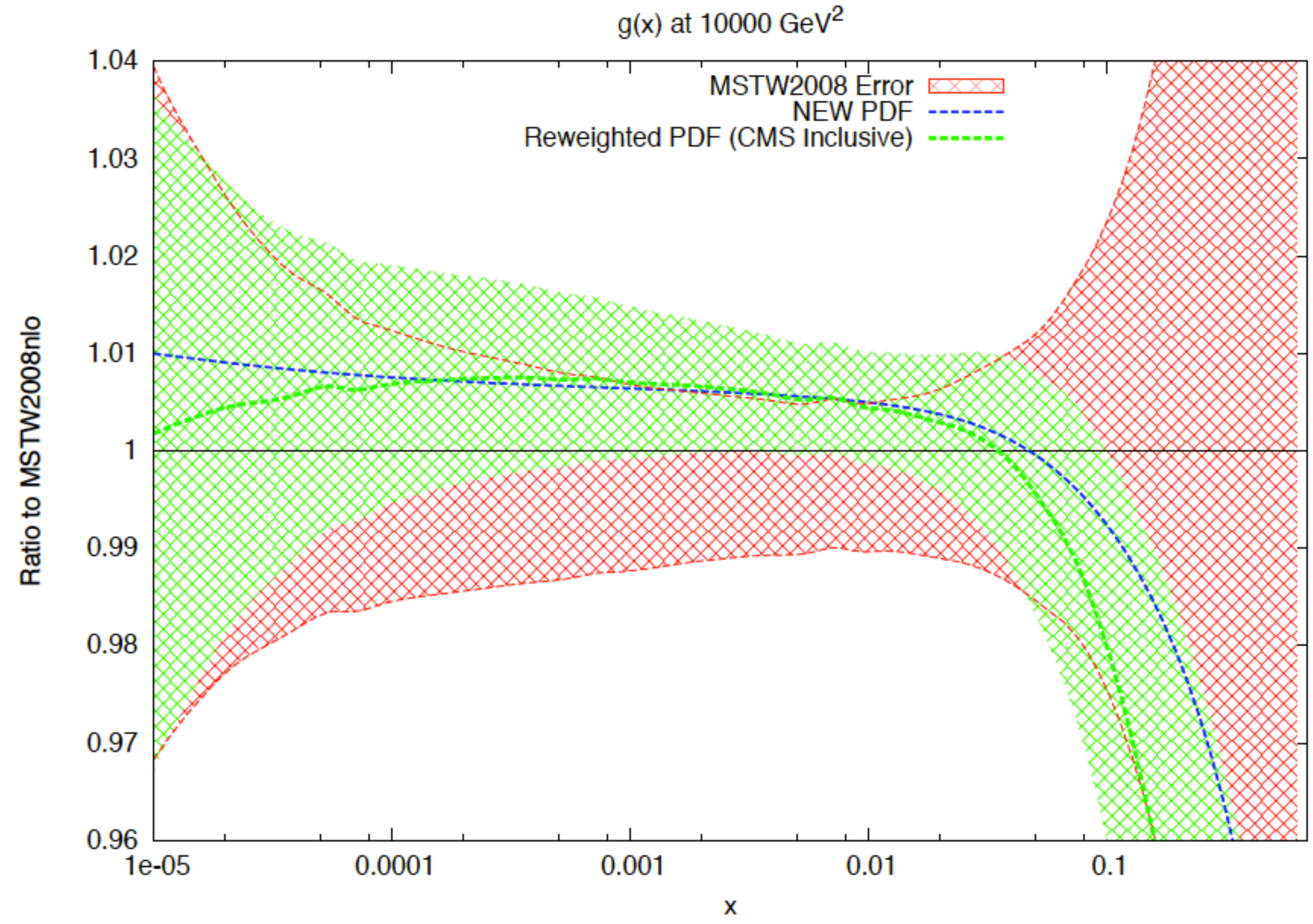}
\includegraphics[width=0.50\textwidth]{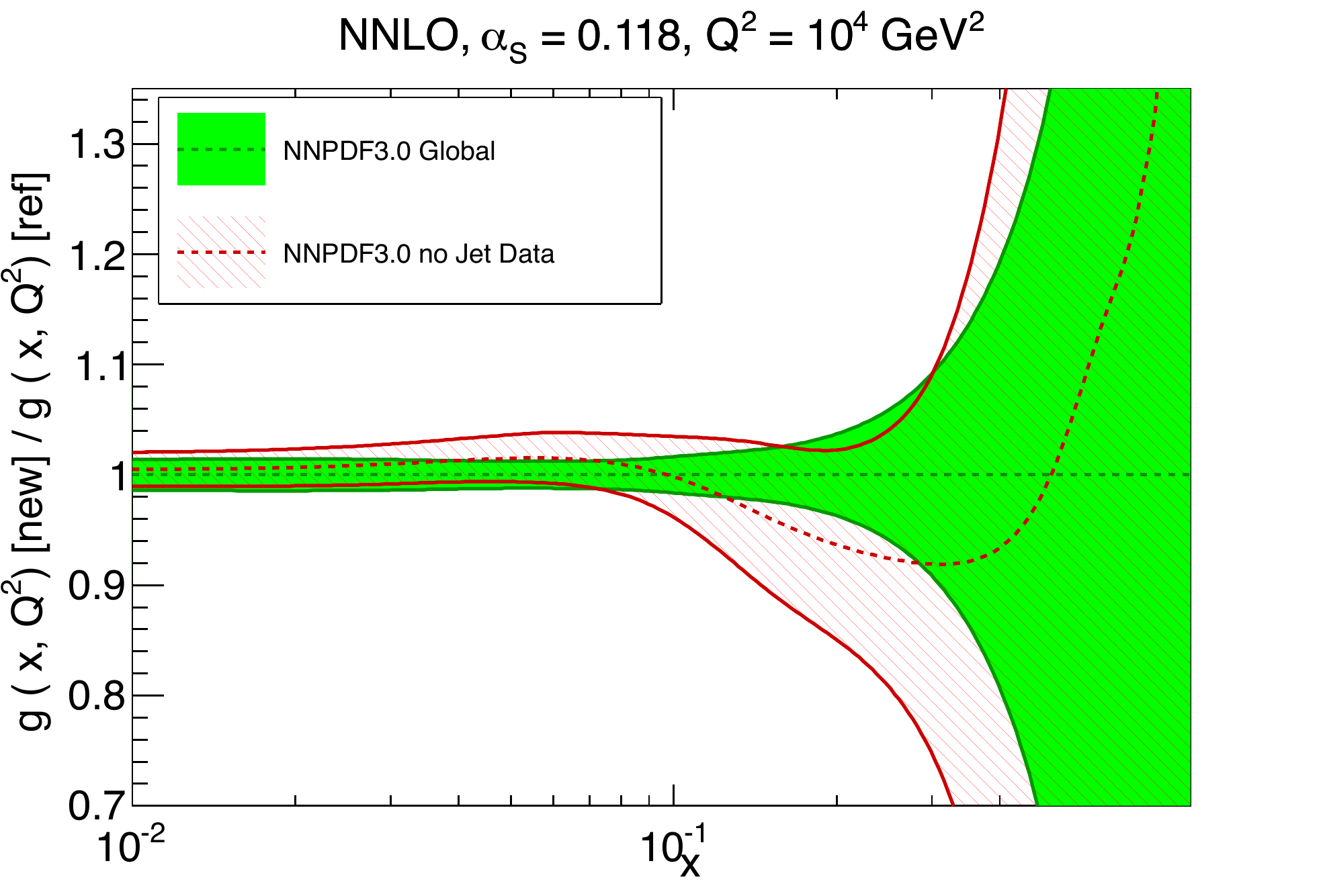}
\end{center}
\vspace{-0.3cm}
\caption{\label{fig:mstwpdf} Left plot: the impact of the CMS
inclusive jet data from the 2011 dataset on the MSTW08 gluon\cite{Watt:2013oha}.
Right plot: the improvement in the large-$x$ gluon PDF
uncertainties thanks to the inclusion of
Tevatron and LHC jet data in the NNPDF3.0 NNLO fit\cite{Ball:2014uwa}.
}
\end{figure}

In addition to these studies
performed by the LHC collaborations, 
most global PDF fitting groups have explored the
impact of the LHC jet data, and some already include them in their most 
recent
releases.
The MSTW collaboration has presented a detailed analysis
of the impact of LHC inclusive jet and dijet data into
their PDFs\cite{Watt:2013oha}, restricted to NLO theory.
As can be seen in Fig.~\ref{fig:mstwpdf}, a reduction of the PDF
uncertainties at large-$x$ from the CMS inclusive jet data is reported,
with the new gluon tending to be softer than the baseline gluon in MSTW08.
Ref.\cite{Watt:2013oha} also finds difficulties in achieving a good
$\chi^2$ for the dijet data, and improving the situation might require
the full NNLO calculation.
See~\cite{Harland-Lang:2014zoa}
for updated studies in the context of the MMHT14 global analysis.

The LHC jet data from ATLAS and CMS is also part of the recent NNPDF3.0 fit,
and is included both in the NLO and NNLO fits, in the latter case 
using the
approximate NNLO calculations as discussed in Ref.\cite{Carrazza:2014hra}
In Fig.~\ref{fig:mstwpdf} a variant of the NNPDF3.0 fit without jet
data is compared to the global fit baseline.
The results that NNPDF3.0 finds
are consistent with those reported by MSTW, in particular for 
substantial error reduction that
LHC data provides.
The CT collaboration has also compared their predictions to LHC jet data
in Ref.\cite{Gao:2013xoa}\, , measurements that will be included in the future CT14 release.
Studies of the impact of hadron collider jet data in the
ABM framework have been reported in Ref.\cite{Alekhin:2012ce}.

%%%%%%%%%%%%%%%%%%%
%%%%%%%%%%%%%%%%%%%%%%%

To conclude this section, it should be clear from the above discussion
that PDF fits with LHC jet data have been so far restricted to
inclusive jet measurements, excluding dijet data.
One important practical reason for this is that typically the correlation
between the two datasets is unknown, and thus dijets are not included
to avoid double counting.\footnote{As mentioned above, this will be no longer
the case with the upcoming ATLAS 2011 jet measurements.}
However, the main reason is that dijet calculations are affected by larger
theoretical uncertainties than inclusive jets, as reflected by the fact 
that the wide choice of scales allowed by dijet kinematics leads to quite
different cross-sections~\cite{Watt:2013oha}.
In this respect, the completion of the NNLO dijet calculation should make
possible to include these data in PDF fits without further problems.

%%%%%%%%%%%%%%%%%%%%%%%%%%%%%%%%%%%%
\section{Determinations of $\alpha_s(M_Z)$ from LHC jets}
%%%%%%%%%%%%%%%%%%%%%%%%%%%%%%%%%%%%

% Use summary alphas plot from CMS

\label{sec:alphas}

Now I turn to review the status of the
determinations of the strong coupling constant 
from LHC data.
The motivation of such determinations is three-fold.
First of all, the uncertainty on the strong coupling
constant  $\alpha_s(M_Z)$ is a substantial contribution
to the total theoretical uncertainty in important LHC processes,
such as Higgs production in gluon fusion\cite{Watt:2011kp,Demartin:2010er,Lai:2010nw}.
Second, accurate  $\alpha_s(M_Z)$ measurements allow to test
the possible unification of the strong, weak and electromagnetic
coupling constants at very high scales~\cite{Dimopoulos:1981yj}.

Finally, the energy reach available at the LHC makes
possible for the first time to
perform direct measurements of $\alpha_s(Q)$ in the TeV scale,
and thus provide model-independent constraints on Beyond
the Standard Model scenarios which are characterized by a different
running of the coupling as compared to QCD above a certain mass
scale, see for example 
Refs.\cite{Becciolini:2014lya,Berger:2010rj,Berger:2004mj}.
Indeed, various studies based on LHC data provide determinations
both of the value of the strong coupling evolved down
to the $Z$ boson mass, $\alpha_s(M_Z)$,  using the RG equations,
as well as direct extractions
of $\alpha_s(Q)$ for $Q\gg M_Z$ that allow to test its running with the scale
$Q$ and to compare with the extrapolations of the SM predictions.

Within the framework of global PDF analyses, jet production
measurements provide a crucial handle, allowing a reliable
extraction of  $\alpha_s(M_Z)$.
In particular, jet measurements from the Tevatron
in global fits play an instrumental role in stabilizing
the gluon, and in turn the closely related value of $\alpha_s(M_Z)$,\footnote{This is
so because from deep-inelastic scattering data only it is difficult
to separate the effects of a change on the gluon PDF from those
of a variation of $\alpha_s$.} 
as discussed for instance in Refs.\cite{Ball:2011us,Thorne:2011kq,Lai:2010nw}.
In particular, the effect of the LHC jet data in the best-fit $\alpha_s(M_Z)$
in the MSTW framework has also been studied\cite{Watt:2013oha}.
In global PDF analyses that include jet data, 
the value of $\alpha_s(M_Z)$ that is extracted  is typically consistent
with the PDG average, which in its most updated version\cite{Agashe:2014kda} reads
\be
\alpha_{\rm PDG}(M_Z) = 0.1185 \pm 0.0006 \, .
\ee
On the other hand, extractions of $\alpha_s(M_Z)$ from PDF
fits without jet data\cite{Alekhin:2009ni}
tend
to be systematically lower.

Concerning direct determinations of $\alpha_s$
from jet measurements at the LHC, a number of analyses have been presented,
mostly by the ATLAS and CMS Collaborations themselves.
Currently all these extractions of
 the strong coupling are restricted to NLO accuracy, and therefore the 
scale uncertainties from perturbative higher orders are the main limiting
factor of the resulting accuracy. 
It is clear that, once the
full NNLO results are available, 
future reanalysis of these  determinations of $\alpha_s$  will substantially reduce the
associated theory uncertainties.
In the meantime, using the approximate NNLO calculations
following the recipe of Ref.\cite{Carrazza:2014hra} might provide
a way forward to reduce these dominant theoretical uncertainties.

The CMS collaboration presented its first extraction of 
 $\alpha_s(M_Z)$  from the measurement of
the ratio of three-jets over two-jet cross-sections\cite{Chatrchyan:2013txa} based on the 7 TeV 2011 data.
This ratio is directly proportional to $\alpha_s(Q)$, where $Q$ is defined
as the average transverse momentum of the two leading jets, that is,
\be
Q \, = \, \la p_{T,1,2}\ra \equiv \frac{p_{T1}+p_{T2}}{2}  \, .
\ee
One important advantage of this ratio is the partial cancellation of experimental
and theoretical systematic uncertainties common to the three-jet
and the two-jet cross-sections.
The result of this analysis is
\be
\alpha_s(M_Z) = 0.1148\,\pm 0.0014\,{\rm (exp)}\,\pm 0.0018\,{\rm (PDF)}
\,\pm 0.0050\,{\rm (th)} \ ,
\ee
using the NNPDF2.1 as input PDF set.\footnote{Results obtained using CT10 and MSTW08 are consistent with those using NNPDF2.1. 
However, if the ABM11 set is used, the extracted value of $\alpha_s(M_Z)$ is
larger, with central value $\alpha_s(M_Z)=0.1214$.}
The precision is thus limited by the QCD scale uncertainties from the NLO
calculation.
In addition to the extraction of $\alpha_s(M_Z)$, separate determinations of $\alpha_s$ in bins
of $\la p_{T1,2}\ra$ are also provided, including the first
direct determination of the strong
coupling constant in the $\sim 1\,{\rm TeV}$ range.
These extractions of $\alpha_s(Q)$ in the TeV scale
provide
model-independent constraints on new physics, which predict a 
different running with the scale as compared
to the SM.
For instance, Ref.\cite{Becciolini:2014lya} uses this CMS measurement of
$R_{3/2}$ to provide
model-independent constraints on new sectors of colored matter.

As mentioned in the previous section, in Refs.\cite{Khachatryan:2014waa} the
CMS collaboration studied the impact on the PDFs of their 2011 7 TeV
inclusive jet production data\cite{Chatrchyan:2012bja}.
In the same analysis, CMS also extracted the
strong coupling, and their
best-fit result turns out to be
\be
\alpha_s(M_Z) = 0.1185 \pm 0.0019\,{(\rm exp)}\pm 0.0028\,{(\rm PDF)}
\pm 0.0004\,{(\rm NP)}\,^{+0.0053}_{-0.0024} \,{(\rm scale)} \ ,
\ee
where again the precision of the determination of the
coupling constant is limited by the unknown higher-order QCD corrections,
followed by the PDF uncertainties.
A related extraction from the three-jet mass distribution, $d\sigma/dM_3$,
has also been reported by CMS.
For this measurement, the preliminary result is
\be
\alpha_s(M_Z)=0.1160\,^{+0.0025}_{-0.0023} \,{(\rm exp,PDF,NP)}\,^{+0.0068}_{-0.0021} \,{(\rm scale)} \ ,
\ee
which is again dominated by scale variations.
All these values of $\alpha_s(M_Z)$ reported by CMS are consistent,
within uncertainties, with the PDG average value.

Turning to the results based on the ATLAS data,
closely related to the CMS analysis based on $R_{3/2}$,
a measurement of $\alpha_s$ from the ratios
of three-jet over two-jet events has been presented by the ATLAS
collaboration\cite{ATLAS-CONF-2013-041} from their 2010 7 TeV
data (that is, based on only $\sim 40$ pb$^{-1}$ of data).
In this analysis, jets are clustered with the anti-$k_T$ algorithm
with $R=0.6$, and only jets with $p_T \ge$ 60 GeV and 
$|y|\le$ 2.8 are included.
Two ratios are defined, the first as a function of the transverse momentum
of the leading jet in the event and the second as a function of all the jets in the event,
as follows:
\be
R_{3/2}(p_T^{\rm lead}) \, \equiv \, \frac{d\sigma_{N_{\rm jet} \ge 3}/dp_T^{\rm lead}}{d\sigma_{N_{\rm jet} \ge 2}/dp_T^{\rm lead}} \,, \qquad
R_{3/2}(p_T^{\rm all\,jets})\, \equiv \,
 \frac{\sum_{i}^{N_{\rm jets}}d\sigma_{N_{\rm jet} \ge 3}/dp_T^{i} }{
\sum_{i}^{N_{\rm jets}}d\sigma_{N_{\rm jet} \ge 2}/dp_T^{i} } \, .
\ee
These two ratios have a direct sensitivity to $\alpha_s$, of
comparable size.
The result of this analysis is
\be
\alpha_s(M_Z) = 0.111 \pm 0.006\,({\rm exp})\,~^{+0.016}_{-0.003}\,
({\rm theory}) \, .
\ee
The uncertainty in the result is again dominated by the 
theoretical uncertainties due to scale variations, followed by the
large experimental uncertainties: recall that this measurement is based 
on the 2010 dataset, with much less statistics than the CMS $R_{3/2}$ measurement.

Still with the 2010 7 TeV data, the ATLAS inclusive jet cross-sections
were used by Malaescu and Starovoitov in Ref.\cite{Malaescu:2012ts} 
to perform an extraction of $\alpha_s(M_Z)$.
The result reads:
\bea
\alpha_s(M_Z) &=& 0.1151\pm 0.0001\,({\rm stat})\,
\pm 0.0047\,({\rm sys})\,
\pm 0.0014\,({\rm p_T\,range})\nonumber \\
&\pm& 0.0060\,({\rm jet\,size})
~^{+0.0044}_{-0.0011}\,({\rm scale})\,
 ~^{+0.0022}_{-0.0015}\,({\rm PDF\,choice}) \\
 &\pm& 0.0010\,({\rm PDF\,eig})\, 
 ~^{+0.0009}_{-0.0034}\,({\rm NP\,corrections}) \nonumber \ .
\eea
In this analysis the dominant systematic uncertainty was found to arise from the
 difference in the results obtained if either data with
jet radius $R=0.4$ or
$R=0.6$ is used, followed by the experimental systematics
(dominated by the jet energy scale) and the unknown perturbative higher orders.

\begin{figure}[h]
\begin{center}
\includegraphics[width=10cm]{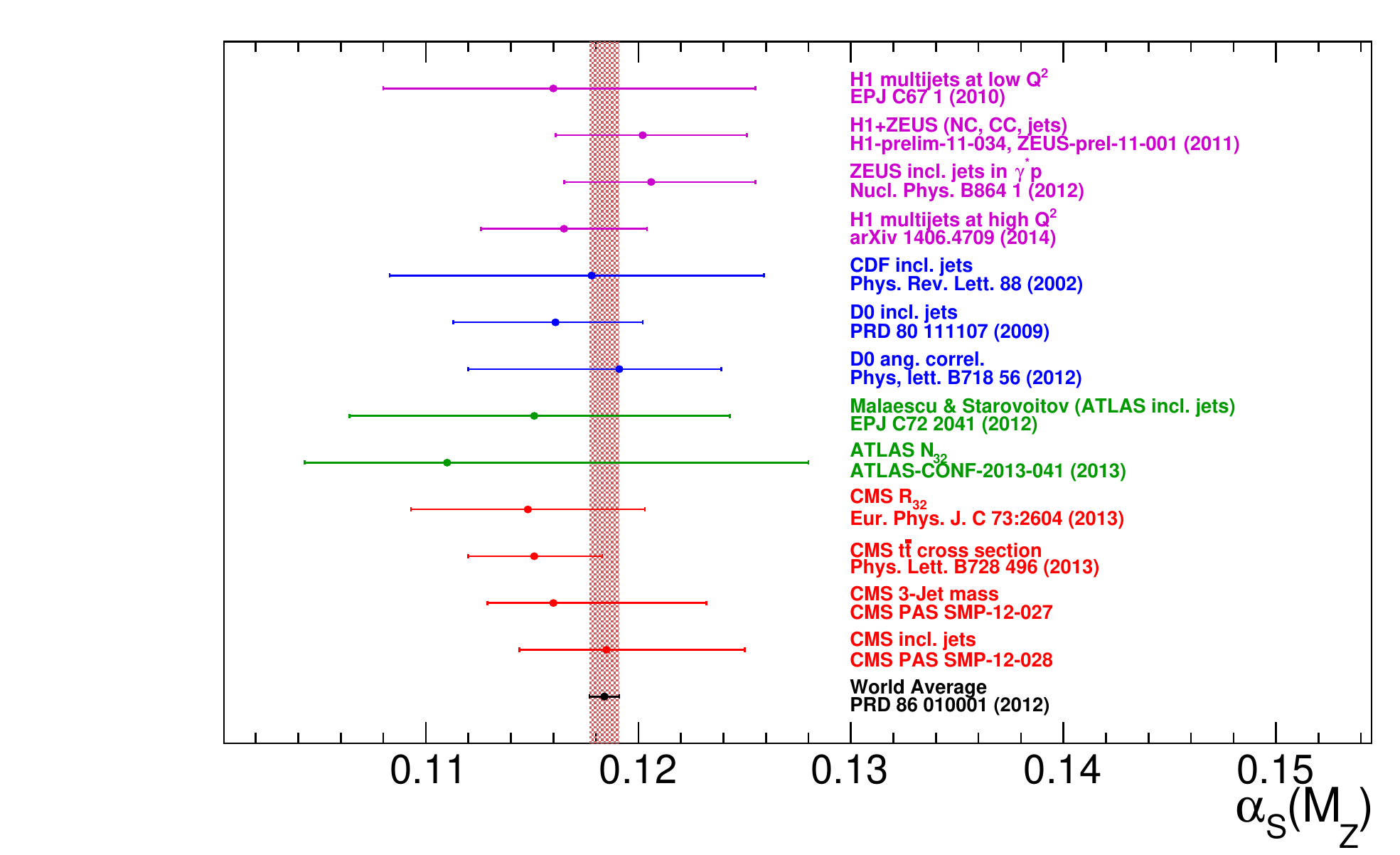}
\end{center}
\vspace{-0.3cm}
\caption{\label{fig:as} Summary of recent determinations
of the strong coupling constant from collider jet data,
together with the PDG world average. 
In each case, the central value and the total uncertainty band are shown.
The plot includes results from HERA and Tevatron together with
the LHC determinations based on
ATLAS and CMS data.
Summary plot taken from the CMS Standard Model Twiki, \url{https://twiki.cern.ch/twiki/bin/view/CMSPublic/PhysicsResultsSMP}.
}
\end{figure}

These results for the various determinations of the strong coupling
from LHC jet data have been collected in Table~\ref{table:alphas}, and summarized
graphically in Fig.~\ref{fig:as},
together with other determinations from collider jet data from
HERA and the Tevatron.
In each case the plot shows the total uncertainty band, and the results of the
individual determinations
are compared with the PDG global average.
The nice consistency of the determinations based on LHC data with the
PDG average is clear from the plot.
For completeness,  Fig.~\ref{fig:as} also includes the results
of other determinations of  $\alpha_s(M_Z)$ from jet data of non-LHC experiments,
like H1, ZEUS, CDF and D0.
It should be emphasized again that all these results are based on
NLO QCD calculations, and therefore one expects substantial
improvements once the same data is reanalyzed using the full
NNLO results for jet production.\footnote{
As an example, the data on hadronic jet shapes
in electron-positron collisions from LEP was reanalyzed in\cite{Dissertori:2007xa} once
the full NNLO calculation became available.}

%%%%%%%%%%%%%%%%%%%
%%%%%%%%%%%%%%%%%%%%%%%

%%%%%%%%%%%%%%%%%%%%%%%%%%%%%%%%%%%%%%%%%%%%%%%%
\begin{table}[t]
\small
\tbl{Summary of direct determinations of the strong coupling
constant $\alpha_s(M_Z)$ from jet measurements at the LHC.
See text for more details of each specific analysis. In various cases,
 $\alpha_s(Q)$ has also been extracted at different scales $Q$ without
assuming the SM running.}
{\begin{tabular}{c|l|l}
\hline
Reference &  Input measurement  & $\alpha_s(M_Z)$ \\[0.2cm]
\hline
\hline
Ref.\cite{Malaescu:2012ts}  & ATLAS 2010 incl jets
&  $0.1151\pm 0.0001\,({\rm stat})\,
\pm 0.0047\,({\rm sys})\,
\pm 0.0014\,({\rm p_T\,range})\,$\\
& & $\pm 0.0060\,({\rm jet\,size})
~^{+0.0044}_{-0.0011}\,({\rm scale})\,
 ~^{+0.0022}_{-0.0015}\,({\rm PDF\,choice})\, $ \\
  & 
&  $\pm 0.0010\,({\rm PDF\,eig})\, 
 ~^{+0.0009}_{-0.0034}\,({\rm NP\,corrections})  $ \\[0.2cm]
\hline
Ref.\cite{ATLAS-CONF-2013-041} &
ATLAS 2010 $R_{3/2}$ &
$0.111 \pm 0.006\,({\rm exp})\,~^{+0.016}_{-0.003}\,
({\rm theory})$\\[0.2cm]
\hline
Ref.\cite{Chatrchyan:2013txa} & CMS 2011  $R_{3/2}$  &
$0.1148\,\pm 0.0014\,{\rm (exp)}\,\pm 0.0018\,{\rm (PDF)}
\,\pm 0.0050\,{\rm (th)}$ \\[0.2cm]
\hline
Ref.\cite{Khachatryan:2014waa}  & CMS 2011 inclusive jets &
$0.1185 \pm 0.0019\,{(\rm exp)}\pm 0.0028\,{(\rm PDF)}$\\
&  & $\pm 0.0004\,{(\rm NP)}^{+0.0053}_{-0.0024} \,{(\rm scale)}$ \\[0.2cm]
\hline
Ref.\cite{CMS:2014mna} &  CMS 2011 three-jet mass & $0.1160^{+0.0025}_{-0.0023} \,{(\rm exp,PDF,NP)}\,^{+0.0068}_{-0.0021} \,{(\rm scale)} $\\[0.2cm]
\hline
\end{tabular}\label{table:alphas}}
\end{table}
%%%%%%%%%%%%%%%%%%%%%%%%%%%%%%%%%%%%%

Finally, it is worth mentioning that at the LHC there are
other processes, other than jet production,  that can be used to extract $\alpha_s(M_Z)$.
One example is provided by the total
top-quark production cross-section\cite{Chatrchyan:2013haa}.
As reported in\cite{Chatrchyan:2013haa}, using the NNLO calculation\cite{Czakon:2013goa}
and NNPDF2.3\cite{Ball:2012wy} as input, the CMS collaboration extracted $\alpha_s(M_Z)$
from their 7 TeV inclusive top quark pair production measurements, finding
the following result:
\bea
\alpha_s(M_Z) &=& 0.1151\,^{+0.0028}_{-0.0027} \,{\rm (tot)} \nonumber\\
&=&  0.1151\,^{+0.0017}_{-0.0018} \,{(\rm exp)}^{+0.0013}_{-0.0011} \,{(\rm PDF)}^{+0.0009}_{-0.0008} \,{(\rm scale)}\,\\
&\pm& 0.0013 \,({\rm m_t^{pole}})\, \pm  0.0008 \,{(\rm E_{\rm LHC})} \nonumber \, . \nonumber
\eea
Note the substantial reduction on scale uncertainties as compared
to the determinations from jet data in Table~\ref{table:alphas}, as a consequence
of the availability of NNLO calculations for this process.
This result is also shown in the summary plot in Fig.~\ref{fig:as}.
Results for the central value of $\alpha_s(M_Z)$ and the corresponding
uncertainties are similar when other PDF sets are used, except for the
ABM11 set which prefers instead a larger value, $\alpha_s(M_Z) =0.1187$.
Note that for ABM11 the central value of $\alpha_s$ from their NNLO
fit is much smaller, $\alpha_s(M_Z)=0.1134$.

%%%%%%%%%%%%%%%%%%%%%%%%%%%%%%%%%%%%
\section{Summary and outlook}
%%%%%%%%%%%%%%%%%%%%%%%%%%%%%%%%%%%%

\label{sec:outlook}

In this review, I have presented an overview of the constraints
on the parton distributions of the proton and on the strong
coupling constant $\alpha_s$ that have been obtained up to now
from jet production measurements at the Large Hadron Collider.
I have summarized various analyses that coincide qualitatively:
LHC jet data provides important constraints on the 
medium and large-$x$
gluon PDF, as well as on the large-$x$ quarks.
I have also discussed recent progress in theoretical
calculations and tools for fitting jet data, and presented
a  possible strategy to include jet data
in NNLO fits based on the use
of approximate NNLO calculations from threshold
resummation, validated with the exact
NNLO calculations.
These studies have so far been restricted to 7 TeV data;
once the full 
Run I data is analyzed one expects more stringent constraints thanks
to the increase in statistics, the corresponding improvement in systematic
uncertainties and the extended lever arm in jet transverse momentum
and dijet invariant mass.
In the medium term, the higher-energy Run II will also provide
a wide range of jet measurements with PDF sensitivity.
In the analysis of Run II data, accounting for NLO electroweak corrections will
be mandatory, since in the TeV region these can be comparable
to QCD effects.
As an illustration, CMS already includes these electroweak effects
in their PDF and $\alpha_s$ study from the 2011 data\cite{Khachatryan:2014waa}, and
the same is true for ATLAS in their dijet measurements from the 2011 run\cite{Aad:2013tea}.

I have then reviewed existing determinations of $\alpha_s(M_Z)$ using
LHC jet data,
and shown that, while experimental uncertainties are typically competitive
with other processes,
the overall precision is degraded by the lack of knowledge of the
full NNLO calculation.
Restricted to NLO theory, all these extractions are so far
consistent within uncertainties with the current
global PDG average.
It is worth emphasizing that thanks to the LHC data
the first direct measurements of $\alpha_s(Q)$ in the TeV region have
been obtained, which provide for the first time important
model-independent tests of the running of the coupling
and search for possible deviations that could arise in New Physics 
scenarios.
It will be important to reanalyze these existing determinations
once the full NNLO result is available, when scale
uncertainties will decrease substantially.
In addition, now that LHC jet data is also being included in most global
PDF fits, it will be  interesting to study what 
is the impact of available and future measurements into the determinations
of $\alpha_s$ in the framework of global PDF fits.
Once the Run II data is available, the kinematical
coverage for high-precision direct extractions of the 
coupling constant will extent deep into the
TeV region,  thus it will provide a unique opportunity to search
and exclude robust  
model-independent constraints on new colored matter sectors.

All in all, the LHC is providing a unique window to study in great
detail the richness of the strong interaction, and jet production
in particular offers a unique opportunity to pin down the parton
distribution functions at large-$x$, a crucial prerequisite for
New Physics searches. 
Remarkably, with the LHC we can directly extract for the first time
$\alpha_s$ in the TeV region, validate if its running with the scale
is consistent with the SM predictions and provide robust model-independent
bounds on new colored sectors.

\section*{Acknowledgments}
I am grateful to Gunther Dissertori for the invitation to write this review
and to Stefano Carrazza, 
Stefano Forte for their suggestions
and to Amanda Cooper-Sarkar, Sasha Glazov, Claire
Gwenlan, Panos Kokkas, Klaus Rabbertz and 
Alessandro Tricoli for their comments on the manuscript and for information
on the ATLAS and CMS jet measurements.
This work has been supported by
 an STFC Rutherford Fellowship ST/K005227/1 and by
the European Research Council with the Starting Grant "PDF4BSM".

%\bibliographystyle{ws-ijmpa}
%\bibliography{JetReview}

\end{document}